\documentclass[12pt]{iopart} % IOP format
\usepackage{graphicx}% include figure files.
\usepackage{enumitem}%
\usepackage{url}
\usepackage[dvipsnames]{xcolor} % adds defined colors. % Not needed in final version.

\begin{document}
	
\title{Resonant Soft X-Ray Scattering on LaPt$_{2}$Si$_{2}$}
	
	\author{Deepak~John~Mukkattukavil}
	\ead{deepak.john\_mukkattukavil@physics.uu.se}
	\address{Department of Physics and Astronomy, Uppsala University, Box 516, SE-751 20 Uppsala, Sweden}
	
	\author{Johan Hellsvik}
	\address{PDC Center for High Performance Computing, KTH Royal Institute of Technology, SE-106 91 Stockholm, Sweden}
	\address{Nordita, KTH Royal Institute of Technology and Stockholm University, Hannes Alfvéns v\"ag 12, SE-106 91 Stockholm, Sweden}

	\author{Anirudha~Ghosh}
	\address{Department of Physics and Astronomy, Uppsala University, Box 516, SE-751 20 Uppsala, Sweden}
	
	\author{Evanthia~Chatzigeorgiou}
	\address{Department of Physics and Astronomy, Uppsala University, Box 516, SE-751 20 Uppsala, Sweden}
	
	\author{Elisabetta~Nocerino}
	\address{Department of Applied Physics, KTH Royal Institute of Technology, SE-106 91 Stockholm, Sweden}
	
    \author{Qisi Wang}
	\address{Physik-Institut, Universit\"at Z\"urich, Winterthurerstrasse 190, CH-8057 Z\"urich, Switzerland}
	
	\author{Karin~von~Arx}
	\address{Physik-Institut, Universit\"at Z\"urich, Winterthurerstrasse 190, CH-8057 Z\"urich, Switzerland}
	\address{Chalmers University of Technology, SE-412 96 G\"oteborg, Sweden}

	\author{Shih-Wen Huang}
	\address{MAX IV Laboratory, Lund University, SE-221 00 Lund, Sweden}
	
	\author{Victor Ekholm}
	\address{MAX IV Laboratory, Lund University, SE-221 00 Lund, Sweden}
	
	\author{Zakir Hossain}
	\address{Department of Physics, Indian Institute of Technology, Kanpur 208016, India}
	
	\author{Arumugum Thamizhavel}
	\address{DCMPMS, Tata Institute of Fundamental Research, Mumbai 400005, India}
	
	\author{Johan Chang}
	\address{Physik-Institut, Universit\"at Z\"urich, Winterthurerstrasse 190, CH-8057 Z\"urich, Switzerland} 
	
	\author{Martin M\aa nsson}
	\address{Department of Applied Physics, KTH Royal Institute of Technology, SE-106 91 Stockholm, Sweden}
	
	\author{Lars Nordstr\"om}
	\address{Department of Physics and Astronomy, Uppsala University, Box 516, SE-751 20 Uppsala, Sweden}
	
	\author{Conny~S\aa the}
	\address{MAX IV Laboratory, Lund University, SE-221 00 Lund, Sweden}
	
	\author{Marcus~Ag\aa ker}
	\address{Department of Physics and Astronomy, Uppsala University, Box 516, SE-751 20 Uppsala, Sweden}
	\address{MAX IV Laboratory, Lund University, SE-221 00 Lund, Sweden}
	
	\author{Jan-Erik~Rubensson}
	\address{Department of Physics and Astronomy, Uppsala University, Box 516, SE-751 20 Uppsala, Sweden}
	
	\author{Yasmine~Sassa}
	\ead{yasmine.sassa@chalmers.se}
	\address{Department of Physics, Chalmers University of Technology, SE-412 96 G\"oteborg, Sweden}
	
\date{\today}
	
\begin{abstract}
X-ray absorption (XAS) and Resonant Inelastic X-ray Scattering (RIXS) spectra of LaPt$_2$Si$_2$ single crystal at the Si $L$
%$_{2,3}$
and La $N$
%$_{4,5}$
edges are presented. The data are interpreted in terms of density functional theory, showing that the Si spectra can be described in terms of Si $s$ and $d$ local partial density of states (LPDOS), and the La spectra are due to quasi-atomic local 4$f$ excitations. Calculations show that Pt $d$-LPDOS dominates the occupied states, and a sharp localized La $f$ state is found in the unoccupied states, in line with the observations. 

%The XAS spectrum shows sharp peaks at 97.1 eV, 101.49 eV and a dominating broad feature with maximum at 117.4 eV which can be assigned to excitation of atomic-like La $4d^{-1}4f$ states. RIXS spectra excited at these resonances show scattering to La $5p^{-1}4f$ final states, and also X-ray Emission (XES) from the dynamically populated $4d^{-1}4f$ $^3D_1$ state to the ground state is observed.
%, including transitions to the bound LS-forbidden $^3P_1$ and $^3D_1$ states and the giant $^1P_1$ resonance, respectively. 
%The Si XES spectrum is stationary on the emission energy scale, and partial Fluorescence Yield (PFY) is used to  construct the Si$L$XAS spectrum. Using density functional theory we show that the Si spectra can be described in terms of Si s and d local partial density of states (LPDOS). While Pt $d$-LPDOS are predicted to dominate the occupied states, a sharp localized La f state is found in the unoccupied states, in line with the observed quasi-atomic excitations. 

%with qualitative correspondence for the partial density of states of Si sites and the XES spectrum.   

%XES spectrum from the sample however shows site selective Si1 and Si2 s orbital excitations. Emission from the dynamically excited $4d^{-1}4f$ $^3D_1$ state to electronic local La $4f^0$ ground state is also present in XES at 101.46 eV. The experiments are complemented with modeling using density functional theory, with qualitative correspondence for the partial density of states of Si sites and the XES spectrum.
%\textcolor{RubineRed}{Mention also PDOS for La?}
\end{abstract}
	
\maketitle
	
\section{\label{sec:Intro}Introduction}
In the last decade, systems combining superconductivity (SC) and charge- and/or spin-density waves (CDW and/or SDW) order have attracted numerous attention. Materials like Fe-based pnictides \cite{Lumsden2010}, transition metal dichalcogenides \cite{Manzeli2017} or cuprate superconductors \cite{Keimer2015} are three examples out of many. The interplay between CDW/SDW and SC remains unclear and subject of vigorous discussions.  %The interplay between   have gained significant interest as the fluctuations associated with CDW and/or SDW are believed to be a key factor for the understanding of SC in these systems.
	
Recently, the quasi-two-dimensional Pt-based rare earth intermetallic material LaPt$_2$Si$_2$ has attracted a lot of attention as it exhibits strong interplay between CDW and SC \cite{Gupta2017}. LaPt$_2$Si$_2$ crystallizes in a CaBe$_2$Ge$_2$-type tetragonal structure (space group P4/nmm, see Fig. \ref{Figure_1}(a)), having a close resemblance to the ThCr$_2$Si$_2$-type structure found in pnictide and heavy fermion systems. The striking difference between these two structures is that the former one lacks inversion symmetry in the crystal, resulting in two non-equivalent layers (Si1–Pt1–Si1) and (Si2–Pt2–Si2) arranged in alternating stacking separated by lanthanum atoms. This special feature in the crystal structure is reminiscent of noncentrosymmetric SC \cite{Gupta2017}, where the lack of inversion symmetry results in nonuniform lattice potential, creating an asymmetric spin orbit coupling.
	
In polycrystalline samples, a first order transition was observed from high temperature tetragonal to low temperature orthorhombic phase, accompanied by a CDW transition around T$_{\rm CDW}$ = 112 K followed by a SC transition at T$_{\rm c}$ = 1.22 K \cite{Gupta2017}. Below T$_{\rm SL}$ = 10 K, superlattice reflections corresponding to (n/3, 0, 0), where n = 1 and 2 were observed, indicating that the unit cell of LaPt$_2$Si$_2$ is tripled in size along the [100] direction at low temperature \cite{Nagano2013}. Theoretical calculations show that the Fermi surface of LaPt$_2$Si$_2$ is of two-dimensional nature \cite{Kim2015} and CDW and SC coexist in the (Si2–Pt2–Si2) layer. %Recent investigations into LaPt$_2$Si$_2$ shows that the SC shows a single-gap $s$-wave model with moderate coupling between electrons and phonons \cite{Nie2021}.
Recent investigations of LaPt$_2$Si$_2$ shows that the SC can be described in a single-gap $s$-wave model with moderate coupling between electrons and phonons \cite{Nie2021}.
%\textcolor{RubineRed}{Next few sentences have been edited, check that the meaning is preserved}
%There are reports of multiple CDW transitions in this material, with a first CDW transition around 175 K at $q_1 = (0.360, 0, 0)$ and $q_2 = (0, 0.360, 0)$ and a second CDW transition around 100 K at $q_1^{\prime} = (0.187, 0.187, 0.500)$ and $q2^{\prime} = (0.187, -0.187, 0.500)$ \cite{Falkowski2020}. This new report \cite{Falkowski2020} differs from the previous report  \cite{Gupta2017} of a single CDW transition in LaPt$_2$Si$_2$.

A CDW transition in this material was found some years ago \cite{Gupta2017}, and multiple CDW transitions were recently observed \cite{Falkowski2020}, with a first CDW transition around 175 K at $q_1 = (0.360, 0, 0)$ and $q_2 = (0, 0.360, 0)$ and a second CDW transition around 100 K at $q_1^{\prime} = (0.187, 0.187, 0.500)$ and $q2^{\prime} = (0.187, -0.187, 0.500)$.

To get further insight into the interactions in this material, we have 
%initially 
performed X-ray absorption (XAS) and resonant inelastic x-ray scattering (RIXS) experiments on a LaPt$_2$Si$_2$ single crystal at the Si $L$
%$_{2,3}$
and La $N$
%$_{4,5}$ 
edges. 

The XAS spectrum shows sharp peaks at 97.1 eV, 101.47 eV and a dominating broad feature with maximum at 117.4 eV which can be assigned to excitation of atomic-like La $4d^{-1}4f$ states. RIXS spectra excited at these resonances show scattering to La $5p^{-1}4f$ final states, and we also observe X-ray Emission (XES) from the dynamically populated $4d^{-1}4f$ $^3D_1$ state to the ground state.
%, including transitions to the bound LS-forbidden $^3P_1$ and $^3D_1$ states and the giant $^1P_1$ resonance, respectively. 
The Si XES spectrum is stationary on the emission energy scale, and Partial Fluorescence Yield (PFY) is used to  construct the Si $L$ XAS spectrum. Using density functional theory we show that the Si spectra can be described in terms of Si $s$ and $d$ local partial density of states (LPDOS). While Pt $d$-LPDOS is predicted to dominate the occupied states, a sharp localized La $f$ state is found in the unoccupied states, in line with the observed quasi-atomic excitations. 

%RIXS has proven to be a powerful tool for studying elementary excitations \cite{Ament2011}. Being a photon-in, photon-out technique, RIXS is bulk-sensitive and naturally includes the charge response of the system \cite{Ament2011}.

%The substantial momentum of X-ray photons makes it possible to observe collective magnetic excitation, charge excitations, orbitons in a wide region of the Brillouin zone (BZ) \cite{Ament2011, Schlappa2012}.\\

\section{\label{sec:exp} Experiment}

\begin{figure*}[ht]
	\begin{center}
		\includegraphics[keepaspectratio=true,width=1\textwidth]{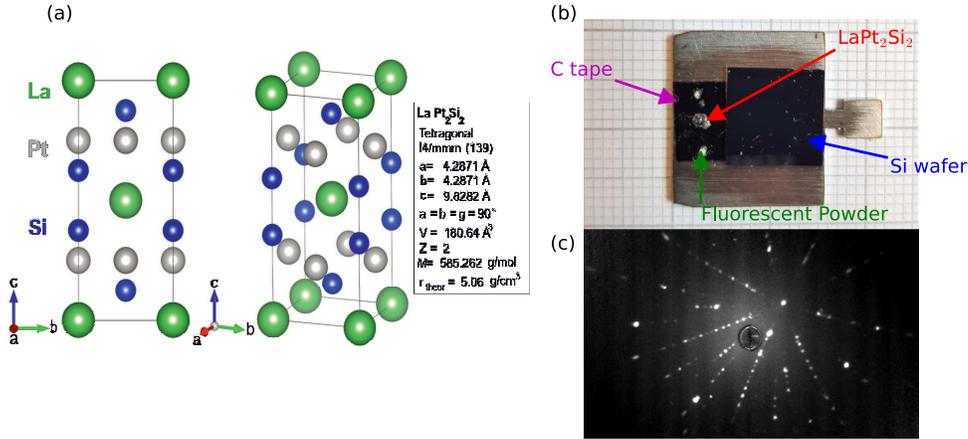}
		\caption{\label{Figure_1}(a) Crystal structure of LaPt$_2$Si$_2$ viewed in two different direction and lattice parameters. (b) LaPt$_2$Si$_2$ crystal mounted on sample plate. Conductive carbon tape was used as adhesive, Fluorescence powder for spotting beam, and a Si wafer was used for calibrating the beamline energy. (c) The Laue diffraction pattern demonstrates the crystallinity of the sample.
		%of the above LaPt$_2$Si$_2$ crystal shows an offset of 14 degrees with (001) direction.
		%{\color{RubineRed}{Use png file instead of large eps file?}}
		}
	\end{center}
\end{figure*}
	
%\emph{Sample synthetization:} 
The single crystalline sample of LaPt$_2$Si$_2$ used in this study was synthesized using the Czochralski pulling method \cite{Gupta2017}. Crystallographic analysis shows that the LaPt$_2$Si$_2$ sample has {{P4/nmm}} space group with the crystallographic parameters given in Fig. \ref{Figure_1}(a). The VESTA software \cite{Momma2011}
was used to generate the crystal structure with crystallographic parameters from Gupta {\it{et al.}} \cite{Gupta2017}.  The size of the sample used in this study 
was $\approx 1$ mm $\times 2$ mm (\ref{Figure_1}(b)). A %Laue 
diffraction pattern, obtained using the Laue backscattering method (Fig. \ref{Figure_1}(c)) shows that the sample is a single crystal.

%\emph{X-ray spectroscopy:} 
XAS and RIXS spectra were measured %during a commissioning beamtime 
at the %RIXS branch of the 
SPECIES beamline \cite{Urpelainen2017,Kokkonen2021} at MAX IV Laboratory in Lund, Sweden. 
%The sample used in this study was a $\approx 1$ mm $\times 2$ mm single crystal as shown in figure \ref{Figure_1}(c). A Laue diffraction pattern on the sample was obtained from using the Laue backscattering method. The Laue diffraction pattern from the LaPt$_2$Si$_2$ sample, figure \ref{Figure_1}(d), shows that the LaPt$_2$Si$_2$ sample is single crystalline in nature with the (001) crystal axis orientation with an offset of $14^{\circ}$. 
%Fluorescence powder was used to visualize the beam spot for alignment purposes, and 
A silicon wafer was attached to the sample holder for energy calibration (Fig. \ref{Figure_1}(b)). Total Electron Yield (TEY) was measured via the drain current, both on the LaPt$_2$Si$_2$ sample, and the silicon reference. Normalization with respect to beamline flux was achieved via the gold-coated refocusing mirror drain current.
% denoted as total electron yield (TEY). %By sweeping the beamline energy from 85-110 eV, 
The TEY measured from the silicon wafer was in good agreement with previous Si L$_{2,3}$ XAS measurements \cite{Kasrai1996}.
%TEY measured from the silicon wafer by sweeping the beamline energy from 85-110eV was in good agreement with previous Si L$_{2,3}$ XAS measurements \cite{Kasrai1996}.
	
XES spectra were measured using the Scientia model XES-350 spectrometer \cite{Nordgren2000}, which is a Rowland spectrometer equipped with three gratings and a micro-channel plate-based delay line detector \cite{Oelsner2001}. For the measurements reported in this article, 
a grating with 3 m radius and 300 l/mm groove density was used. %The spectrometer is a Rowland spectrometer equipped with a micro-channel plate-based delay line detector \cite{Oelsner2001}. 
Energy calibration of the spectrometer relative to the monochromator was done using the elastic scattering of the incident photon beam, and the monochromator energy scale was set by the Si XAS reference \cite{Kasrai1996}.  
%on the Silicon wafer for discrete intervals of beamline energy from 98 to 130 eV. 
The overall resolution was estimated from the full width at half maximum of the elastic scattering peak to be 270 meV at 108 eV incident photon energy. All RIXS measurements were carried out in the horizontal plane at 90$^{\circ}$ scattering angle, and the incident radiation was linearly polarized in the horizontal direction.
	
%The second set of XES measurements were done on the sample at the same beamline with the same grating under grazing incidence geometry with an incidence angle of 20$^o$ with a resolution of 230 meV at 95eV incident photon energy.

%{\textcolor{blue}{The only info in that paragraph is that we have 20$^o$ angle of incidence --- what is first or second set is not really relevant, but it is not sais anywhere wht the angle was in the first set ... I think close to 45$^o$}

\section{\label{sec:theory} Theory}

Density functional theory (DFT) calculations of the electronic structure of LaPt$_2$Si$_2$ have been performed with the Elk code \cite{elk} which implements the full-potential augmented plane waves and local orbitals method (FP-APW+lo) \cite{Singh2006ppa}.
In tetragonal CaBe$_2$Ce$_2$ structure (space group $P4/nmm$, setting 2), the primitive cell of LaPt$_2$Si$_2$ contains 10 atoms. Calculations have been performed for this structure, using the lattice parameters reported by Gupta \emph{et al.} \cite{Gupta2016}. 
We used the generalized gradient approximation (GGA) as parametrized in the PBE functional \cite{Perdew1996} for
nonmagnetic, collinear spin-polarized, and noncollinear magnetic including spin-orbit coupling (SOC) calculations, using a maximum angular momentum $l=8$, and a $\Gamma$-centered $\textbf{\emph{k}}$-point mesh of size $16\times16\times16$. 

%The Elk calculations were performed using the generalized gradient approximation (GGA) as parametrized in the PBE functional \cite{Perdew1996}.

%Calculations have been performed as nonmagnetic, collinear magnetic and noncollinear magnetic including spin-orbit coupling (SOC). In Sec. \ref{sec:results} will be mainly discussed the results from the calculations with SOC.

%In tetragonal CaBe$_2$Ce$_2$ structure (space group $P4/nmm$, setting 2), the primitive cell of LaPt$_2$Si$_2$ contains 10 atoms. Calculations have been performed for this structure, using the lattice parameters reported by Gupta \emph{et al.} \cite{Gupta2016}. The atomic positions were relaxed within the Elk calculations until the residual forces were less than {\color{RubineRed}{SPECIFY VALUE. Need convert Elk Hartree and Bohr radius to eV and \AA}}.

\section{\label{sec:results}Results and Discussion}
\begin{figure}
\begin{center}
\includegraphics[width=0.60\textwidth]{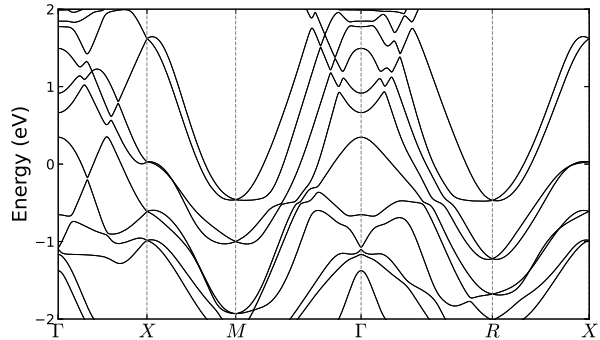}
\caption{\label{fig:ElkSOCbands}Electronic band structure in Elk PBE SOC calculation. Nearly degenerate bands are split by the spin-orbit coupling.}
\end{center}
\end{figure}

\begin{figure}
\includegraphics[width=0.6\textwidth]{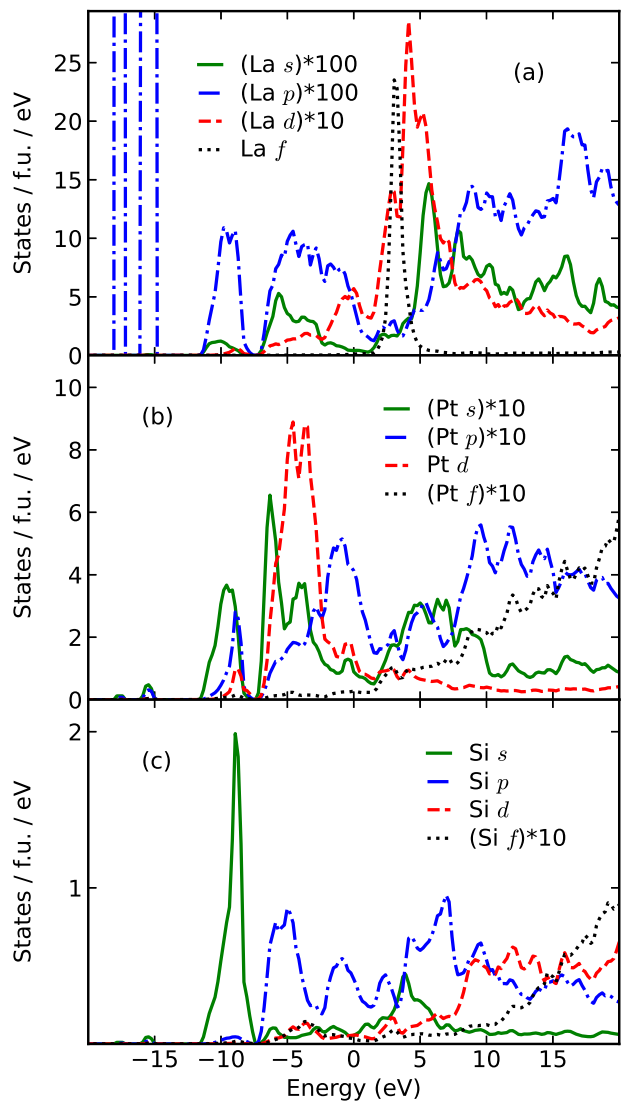}
\caption{\label{fig:ElkSOCDOS}Partial electronic DOS in Elk PBE SOC calculation for (a) La, (b) Pt, and (c) Si. In order to accommodate the data in the panels, some of the components have been multiplied with a factor of 10 or 100.}
\end{figure}

In the Elk PBE collinear spin-polarized calculation, the electron band structures for majority and minority spins comes out as  degenerate, from which we infer that LaPt$_2$Si$_2$ is a non-magnetic material.
The electronic band structure in Elk PBE SOC calculation is shown in Fig. \ref{fig:ElkSOCbands}, with the associated total DOS, and projected partial DOS for La, Pt, and Si states shown in Fig. \ref{fig:ElkSOCDOS}. At the La site a sharp La $f$-LPDOS peak dominates the unoccupied states, and two sharp peaks in the La $p$-LPDOS are found below -15 eV in the occupied states (Fig. \ref{fig:ElkSOCDOS}a). It can be concluded that the La states are localized and do not contribute much to the band formation.
The occupied states are dominated by Pt $d$-LPDOS with a maximum around -4 eV (Fig. \ref{fig:ElkSOCDOS}b) whereas $p$-symmetry gives the largest contribution to the electronic structure at the Si sites (Fig. \ref{fig:ElkSOCDOS}c), apart from a sharp peak in the Si $s$-LPDOS at around -10 eV.
By virtue of the dipole selection rules, X-ray spectra at the Si $L$ edges primarily probe the Si $(s+d)$-LPDOS, and reflect the localized nature of the La $f$-excitations at the La $N$ edges. 

\begin{figure*}[ht]
	\begin{center}
		\includegraphics[keepaspectratio=true,width=1\textwidth]{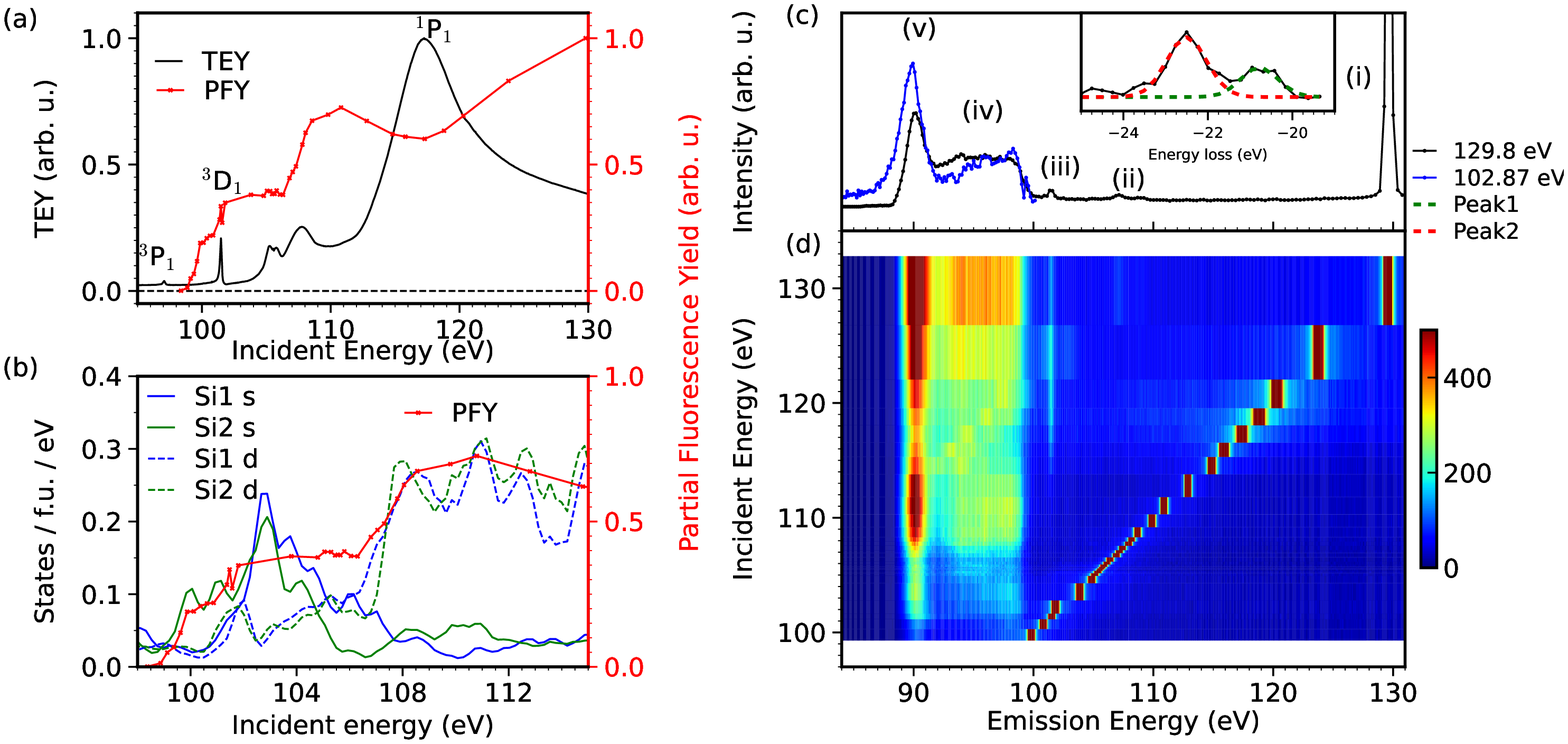}
		\caption{\label{Figure_2}(a) TEY (black) measured using drain current from sample and PFY (red) for emission in the 85-98 eV range, with the dispersing feature (ii) subtracted. (b) Zoomed-in PFY for emission in the 85-98 eV range (red) with DOS of Si1 $s$ (blue), Si2 $s$ (green), Si1 $d$ (dash blue), Si2 $d$ (dash green) above Fermi energy. (c) An overview XES spectrum from LaPt$_2$Si$_2$ at incident photon energy of 129.8 eV, covering emission energies up to 130 eV (black). In the corresponding detector position, the low-energy part of the spectrum is not fully covered. 
		%{\color{RubineRed}{Should it here read 102.7 eV or 102.87 eV?}}
		The spectrum showing the full low-energy range is measured in a separate measurement at 102.87 eV 
		%102.7 eV 
		excitation energy (blue). The two spectra are normalized in the 95-100 eV range. The emission features are denoted: (i) elastic peak, (ii) resonant scattering to La 5$p^{-1}$4$f$ final states, which is also shown in the inset on the energy-loss scale, (iii) emission from the dynamically populated La 4$d^{-1}$4$f$ $^3D_1$ state, and (iv,v) stationary feature primarily associated with Si $L_{2,3}$ emission. (d) XES map. Like in the overview spectrum in panel (c) the attenuation of the XES emission at the lowest energies is an experimental artefact due to the drop in sensitivity at this detector position.
		}
	\end{center}
\end{figure*}

Figure \ref{Figure_2}(a) shows TEY obtained by the drain current from LaPt$_2$Si$_2$ as a function of incident photon energy. There are two sharp peaks at 97.1 eV and 101.47 eV, and a dominating broad feature with maximum at 117.4 eV, %(Fig. \ref{Figure_2}(a)), 
which can be assigned to excitation of atomic-like La $4d^{-1}4f$ states, including transitions to the bound LS-forbidden $^3P_1$ and $^3D_1$ states, and the giant $^1P_1$ resonance, respectively. Similar multiplet structures are typically observed in La compounds \cite{Mueller1995, Ichikawa1986, Rubensson1997, Suljoti2009}, as a consequence of the atomic-like localized nature of the $4f$ electrons.

Notably, the TEY spectrum does not capture the Si $L$ edge XAS of LaPt$_2$Si$_2$, expected to appear just below 100 eV. This observation demonstrates that the La $4d$ absorption cross-section totally dominates over Si $2p$ absorption. Instead, we observe TEY peaks between 104 and 108 eV (Fig. \ref{Figure_2}(a)) that match previous XAS spectra of SiO$_2$ \cite{Kasrai1996}. We attribute these peaks to an oxide layer formed on the surface of LaPt$_2$Si$_2$ crystal.

%The observation directly demonstrates that the La $4f$ level is unoccupied in the LaPt$_2$Si$_2$ ground state.

%The TEY peaks (figure \ref{Figure_2}(a)) between 104 and 108 eV match previous XAS spectra of SiO$_2$ \cite{Kasrai1996}, and can be assigned to SiO$_2$ formed on the surface of LaPt$_2$Si$_2$ crystal. 
%There are two sharp peaks at 97.1 eV, 101.47 eV and a dominating broad feature with maximum at 117.4 eV (figure \ref{Figure_2}(a)), which can be assigned to excitation of atomic-like La $4d^{-1}4f$ states, including transitions to the bound LS-forbidden $^3P_1$ and $^3D_1$ states and the  giant $^1P_1$ resonance, respectively.

%Conventional TEY does not capture the Si L edge XAS of LaPt$_2$Si$_2$, even though it is expected in the energy range above 98.5 eV. This demonstrates that the La $4d$ absorption cross-section here totally dominates over Si $2p$ absorption. 
%The spectral features in the 104-110 eV range signify of SiO$_2$ \cite{Kasrai1996}, which we attribute to an oxide surface layer.  -------we have said that already.

To address the Si $L$ absorption spectrum of LaPt$_2$Si$_2$  we first turn our attention to the XES map in Fig. \ref{Figure_2}(d). A complex pattern is observed, where the main intensity is stationary between 87 eV and 98 eV emission energy, primarily changing its integrated intensity with excitation energy. There is also an elastic peak for which the emission energy equals the excitation energy. A weaker structure disperses at constant 20-24 eV energy loss, and a sharp emission feature is observed at 101.46 eV constant emission energy. The latter is resonantly excited at incident photon energies around 120 eV. These features are denoted in the XES spectrum excited at 129.8 eV, shown in Fig. \ref{Figure_2}(c):
	
\begin{enumerate}[label=(\roman*)]
		
	\item Elastic scattering, coinciding with the excitation energy.
		
	\item A constant-energy-loss feature which we assign to the scattering to 5$p^{-1}$4$f$ final states, reached over the 4$d^{-1}$4$f$ resonances. We tentatively assign the two peaks at 20.8 and 22.5 eV energy loss to final triplet states and the $^1D_2$ state, respectively, following the analysis of Suljoti {\it{et al.}} \cite{Suljoti2009} and Miyahara {\it{et al.}} \cite{Miyahara1986}.
		
	\item A sharp peak at 101.46 eV emission energy assigned to emission from the dynamically excited $4d^{-1}4f$ $^3D_1$ state to the La $4f^0$ ground state.
		
	\item A broad feature at constant emission energy, which we assign to principally Si $L_{2,3}$ emission, \textit{i.e.}, is associated with electrons from the valence band filling Si $2p$ vacancies.
		
	\item Si $L_{2,3}$ emission as in (iv), but as we see, the excitation energy dependence of the two features is different.
		
\end{enumerate}

%The cutoff for peak (v) in figure \ref{Figure_2}(c) at 129.8 eV excitation energy is because the detector ends around 87 eV. This part of the spectrum was re-measured, and the actual width of peak (v) is shown by the blue curve 102.87 excitation energy in figure \ref{Figure_2}(c). ----- Ruben thinks it is better to put this in fIg. caption.
	
%\subsection{Partial Fluorescence Yield}
The red curve in Fig. \ref{Figure_2}(a and b) is the Partial Fluorescence Yield (PFY), constructed to emphasize Si $2p$ excitations. As Si $L_{2,3}$ emission dominates features (iv) and (v), intensity in the corresponding 85-98 eV emission energy range was integrated to construct the PFY spectrum, while intensity from the dispersing La feature (ii), which contributes in the 110-120 eV excitation-energy range (Fig. \ref{Figure_2}c inset) was subtracted.
	
The Si $L_{2,3}$ edge shown in the PFY spectrum is found at 99.5 eV, and there is a sharp second intensity increase at 101.4 eV, almost coinciding with the La $^3D_1$ resonance, and another intensity increase around 107 eV.

There is only a faint $(<10\%)$ structure in PFY at the sharp SiO$_2$ excitations around 104 eV which can be attributed to the surface oxide. For oxidized silicon surfaces, a similar observation has been made for oxide layers of around 7 nm thickness \cite{Kasrai1996}. While the fluorescence yield sampling depth in silicon is estimated to be 70 nm, \cite{Kasrai1996}, it is similar in this compound except for the region of the La giant resonance $^1P_1$ where the sampling depth is more than three-fold reduced \cite{Henke1993}. We can therefore estimate the oxide layer also for LaPt$_2$Si$_2$ to be in the 7 nm range.
	
While the Si $L_{2,3}$ edge PFY spectrum gives information about the bulk material below the surface oxide layer, it does not directly represent the cross-section for Si $2p$ excitations. The PFY minimum at 117.3 eV coincides with the maximum of the La giant resonance. At these energies, La $4d$ absorption dominates the XAS spectrum of LaPt$_2$Si$_2$. As this process competes with Si $2p$ absorption, and since La $4d$ holes do not emit appreciably in the chosen emission energy range, the observed broad dip in the PFY is expected. The mechanism is the same as exploited in the inverse partial yield method \cite{Achkar2011}.

In Fig. \ref{Figure_2}(b) we compare the the PFY with the Si $s$ and $d$ LPDOS, separated into contributions from the Si1 and Si2 sites (see Fig.~\ref{Figure_1} for notation). Significant differences between the two sites are predicted, e.g., the absorption close to the edge is primarily of Si2 $s$ character. The PFY increase around 101.4 eV coincides with increasing Si1 $s$-LPDOS, while the 107 eV increase matches the increase in Si $d$-LPDOS. We assign the main PFY features accordingly. 
%although the theoretical LPDOS and the PFY spectrum do not quantitatively agree. 

%$s$ and $d$ states above Fermi energy (Figure \ref{Figure_2}(b)) shows that the significant contribution for PFY is from Si $s$ and $d$ states. The initial increase in PFY is due to Si2 s states being populated. For incident photon energy above 101.5eV, Si1 s states contribute to PFY. Above 106 eV incident photon energy Si1 and Si2 $d$ states contribute to PFY. Thus the DOS above the Fermi energy of Si $s$ and $d$ states {\color{RubineRed}{Rephrase, we cannot say too much about the quantitative agreement}} matches with the PFY.
	
\subsection{Elastic Peak (i)}
In general, the elastic peak (i) can have contributions from both diffuse reflection and the resonant absorption-emission process. Emission from the giant resonance in lanthanum compounds in low-energy electron excited spectra seemingly coincides with the absorption \cite{Ichikawa1986}, demonstrating that a captured electron is sufficiently localized to recombine with the $4d$ hole. The giant resonance is also observed in reflection measurements \cite{Mueller1995}. Considering the complex index of refraction, $n+ik$, it has been noted that if $(n-1)^2<<k^2<<(n+1)^2$ the normal-incidence reflectivity varies as the square of the absorption cross section. While this is often the case in this energy range \cite{Obrien1991}, this approximation does not hold at the giant resonance, for which tabulated optical constants \cite{Henke1993} imply a high-energy shift of the reflectivity peak relative to the peak in the absorption cross section. This predicted shift is consistent with the observed difference in peak positions when comparing the excitation-energy dependence of the elastic peak intensity and the TEY spectrum (Fig. \ref{Figure_3}). While the curves show similarities, there are also significant differences, the TEY spectrum peaks at 117.5 eV, and the maximum in the elastic-peak intensity is around 119 eV. We therefore conclude that the elastic peak is dominated by diffuse reflection in the region of the giant resonance.
	
\subsection{Dispersing Feature (ii)}
The dispersing double-peak feature (ii), assigned to lanthanum $\textrm{ground state}\rightarrow 4d^{-1}4f\rightarrow 5p^{-1}4f$ scattering is shown in the inset of Fig. \ref{Figure_2}(c). The two peaks are fitted with Gaussians at 22.5 and 20.7 eV  energy loss, with full width at half maximum (FWHM) of 1.1 eV and 1.0 eV, respectively. The energy positions are close to the corresponding transitions in ionic lanthanum compounds \cite{Rubensson1997, Suljoti2009}, facilitating the assignment of the high-energy loss peak to the $^1D_2$ final state, and the low-energy loss peak to ``spin-flip'' $^3D_J$ states. The intensity ratio of the two peaks also coincides with the earlier studies, again demonstrating the atomic-like nature of the scattering at the La sites.
	
In the ionic compounds the $5p^{-1}4f$ final states are situated in the band gap, whereas in the present case they are found in the conduction band far above the $5p^{-1}$ ionization thresholds. Therefore, delocalization via interaction with the continuum is allowed, just like in the case of the giant $4d^{-1}4f$ $^1P_1$ resonance.
%, but unlike the sharp triplet resonances. 
We tentatively attribute the additional width to tunneling of the excited electron through a barrier, analogously to the case of the singlet $4d^{-1}4f$ state. With this interpretation the increased width implies a reduction of the lifetime of the $5p^{-1}4f$ final states to less than 1 fs due to this additional decay channel.

%{\color{RubineRed}{
%The lifetime width of the sharp $LS$ forbidden $4d^{-1}4f$ peaks in absorption is larger in lanthanum metal than in ionic LaF$_3$. This is attributed the larger local electron density in the metal, which increases the electronic decay rate, relative to the ionic compound \cite{Miyahara1986}.}}

%{\color{blue}{
%	Do you think that this is relevant here? This is tentative and speculative, but I believe that if you are above the ionization limit, the tunneling mechanism makes delocalization and width, whereas below the ionization limit it is rather Auger-type electron reorganization.  ????
	
%but in any case, one should try also a Lorentzian fit... i believe it would work well too..
%}}

\subsection{Emission at 101.46 eV (iii)}

Within the measurement accuracy, the energy of the stationary feature (iii) (Fig.~\ref{Figure_2}c),  101.46 ($\pm0.09$)~eV, matches the energy of the La $4d^{-1}4f$ $^3D_1$ resonance (101.47 eV). Therefore, it can be assigned to transitions from this $^3D_1$ state to the lanthanum ground state. 

The mechanism for off-resonance population of $4d^{-1}4f$ states in lanthanum compounds has been addressed earlier \cite{Mueller1995, Ichikawa1986}, and the associated population of La $3d^{-1}4f$ states have been termed ``shake-down'' \cite{Okusawa1987}. 

The transition energy is below the
$4d$ binding energies \cite{Kowalczyk1974}, i.e., the $4d$ hole pulls down the $4f$ orbitals below threshold. Consequently, 
the $4d$ holes are screened as an electron from the valence band fills the $4f$ level, forming
$4d^{-1}4fV^{-1}$ states, where $V^{-1}$ denotes a hole in the valence band. If this hole in the valence band delocalizes prior to the decay, the transition mirrors the $^3D_1$ resonance.

\begin{figure}[ht]
	\begin{center}
		\includegraphics[keepaspectratio=true,width=0.5\textwidth]{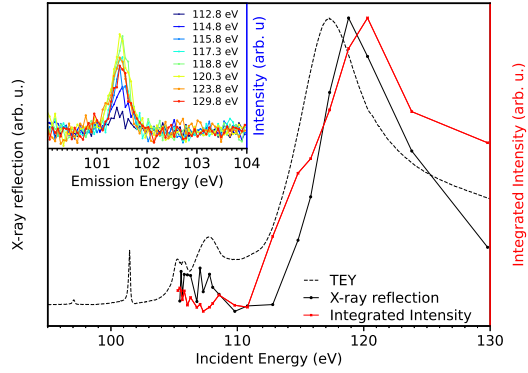}
	    \caption{\label{Figure_3} Elastic peak (i) intensity, dominated by diffuse reflectivity (black), and intensity of the stationary feature ((iii) see inset), associated with La $4d^{-1}4f$ $^3D_1$  $ \rightarrow 4f^0$ transition  (red) compared to the TEY spectrum (dashed black). %When overlapping the stationary peak, the dispersive feature (ii) has been subtracted. 
	    
	    %X-ray reflection (black) obtained from integrated area at energy gain (1-5eV) side of elastic peak indicates. Integrated area of emission line (red) at 101.46 eV emission energy. Overlapping 5$p^{-1}$4$f$ transition features are subtracted in integrated area of emission line (red) curve. The emission line has no intensity below 111 eV. TEY measurement of LPS shown in dotted line for comparison. XES spectra between 100-104 eV above 110eV incident photon energy is shown in inset.
    	%{\color{RubineRed}{User larger font size for inset}}
    	%{\color{RubineRed}{Adjust position of legends}}
	    }
	\end{center}
\end{figure}

In Fig. \ref{Figure_3} we see that the intensity of the $^3D_1$ emission to a large extent follows the absorption cross section, which is in line with the notion that the cross section in the range is totally dominated by La $4d$ excitations. However we find the maximum in the $^3D_1$ yield slightly above 120 eV, and more than 2.5 eV above the absorption maximum. This shift can be understood by considering the localization of the initially excited electron. It is well-known that the probability that this electron tunnels through the potential barrier prior to the core hole decay increases with excitation energy over the giant resonance \cite{Sairanen1991, Rubensson1997}. As long as the initial electron remains local it contributes to the screening of the $4d$ hole and hampers alternative screening mechanisms. Only when it delocalizes the screening from the valence band to the triplet coupled $4d^{-1}4fV^{-1}$ states becomes important. In this way the $^3D_1$ yield in this range reflects a dynamical process where the $4d^{-1}4f$ $^1P_1$ excitation is followed by tunneling of the $4f$ electron, and subsequent screening from the valence band to triplet coupled $4d^{-1}4fV^{-1}$ states. In this sense the $^3D_1$ yield monitors the cross section for excitation of unscreened La $4d$ holes.\\
	
\begin{figure}[ht]
	\begin{center}
		\includegraphics[keepaspectratio=true,width=0.5\textwidth]{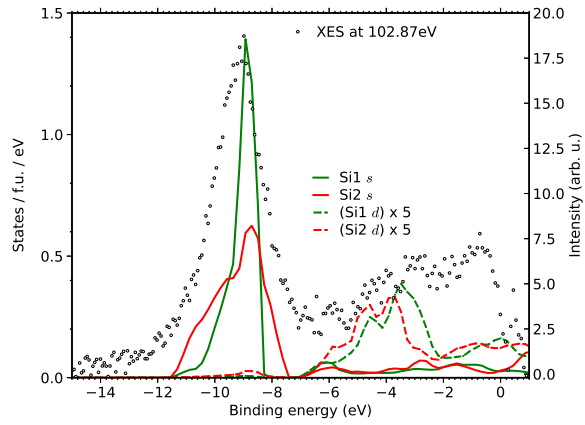}
	    \caption{\label{Figure_4} XES spectrum excited at 102.87 eV, shifted to the binding energy scale by subtracting 98.7 eV (black dots) is compared to the Si1 (full green line) and Si2 (red full line) $s$-LPDOS, and the  Si1 (dashed green line) and Si2 (dashed red line) $d$-LPDOS, below the Fermi level.
	    %Solid green and blue lines are Lorentzian bordered Si1+Si2 s and Si1+Si2 d states respectively. Green dashed lines is sum of Si1+Si2 s and d states.
	    %\caption{\label{Figure_4}(a) Si1 and Si2 s and d states below Fermi energy calculated using DFT. (b) XES spectrum at 129.8eV incident photon energy (black circles). Solid green and blue lines are Lorentzian bordered Si1+Si2 s and Si1+Si2 d states respectively. Green dashed lines is sum of Si1+Si2 s and d states.
	    }
	\end{center}
\end{figure}

\subsection{Si $L$ emission (iv and v)}

\subsubsection{LPDOS and XES}

The emission shown in Fig.~\ref{Figure_4}, excited at 102.87 eV, is shifted to the binding energy scale by subtracting 98.97 eV, as determined by the apparent valence band edge. From the edge, the spectrum shows a plateau down to around -6 eV, followed by a peak with a maximum around -9.5 eV.  The experimental results are compared to the Si $s$ and $d$-LPDOS (Fig.~\ref{Figure_4}) at the two crystallographic sites, Si1 and Si2, as denoted in Fig. \ref{Figure_1}. The low-energy peak can be unambiguously assigned to transitions from states of Si $s$ character as the Si $s$-LPDOS dominates in this region. In the predicted LPDOS there is a marked difference between the two sites. Whereas the Si1 $s$-LPDOS has a rather sharp peak at -9 eV, the Si2 $s$-LPDOS is more smeared out. Due to the broadening mechanisms, especially the lifetime broadening of final-hole states far from the Fermi level \cite{Knapp1979} we do not expect this difference to be observable. The FWHM of the experimental peak is around 2 eV, while a Lorentzian broadening of 1 eV is applied to the theoretical LPDOS.

The Si $s$-LPDOS is not sufficient to capture the plateau in XES spectrum between the valence band edge and -6 eV, as density relative to the -9 eV peak is small in this region. Assuming that the Si $d$-LPDOS gives a factor of 5 larger contribution to the XES intensity than the Si $s$-LPDOS, the intensity in this region can be partly explained (Fig.~\ref{Figure_4}). Since the Si $2p-d$ radial overlap typically is larger than the $2p-s$ overlap, Si $d$-states are expected to contribute more to Si $L$ emission than Si $s$-states \cite{Yorikawa2011}, but a factor of 5 is unusually large. We also note that upscaled Si $d$-LPDOS introduces spectral structure that is not observed in the experimental spectrum. 

%bFor other broad-band compounds it has indeed been seen that Si $d$-LPDOS gives a larger contribution to the emission than the Si $s$-LPDOS because the Si 2p-d radial overlap is larger than the 2p-s overlap \cite{Yorikawa2011},

%The sum of partial DOS of Si1 and Si2 does not really capture the broad feature in XES between zero and -6 eV. The theoretical DOS calculations of LaPt$_2$Si$_2$ thus does not capture the full intensity of Si $L_{2,3}$ edge emission, but we can distinguish regions of XES spectrum contributing from either the $s$ or $d$ states of Si.
%Si1 $s$ states overlap most of the XES peak around -9 eV represented by solid green line in figure \ref{Figure_4}(b). Sum of partial DOS of Si1 and Si2 does not really capture the broad feature in XES between zero and -6eV, which is shown as solid blue curve in figure \ref{Figure_4}(b). Theoretical DOS calculations of LaPt$_2$Si$_2$ thus does not capture the full intensity of Si $L_{2,3}$ edge emission, but we can distinguish regions of XES spectrum contributing from either the $s$ or $d$ states of Si.
	
\begin{figure}[ht]
	\begin{center}
	    \includegraphics[keepaspectratio=true,width=0.5\textwidth]{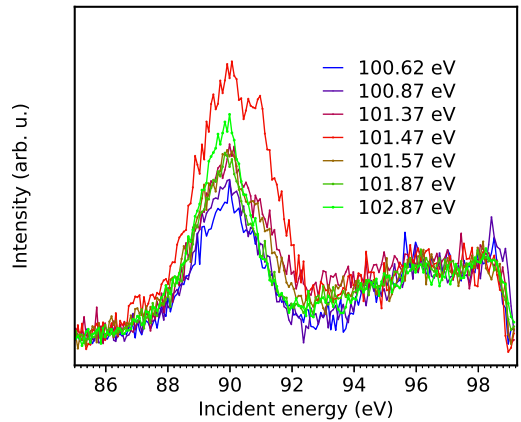}
		\caption{\label{Figure_5} XES spectra normalized at 98 eV, excited in the region of La $4d^{-1}4f$ $^3D_1$ resonance.}
	\end{center}
\end{figure}

\subsubsection{Excitation-Energy Dependence}
	
The spectral changes as the excitation energy varies across the La $^3D_1$ absorption peak are shown in Fig. \ref{Figure_5}. We find that the spectra normalized at 98 eV, show a steep resonant behavior where the intensity of the peak at 90 eV (-9 eV peak in Fig. \ref{Figure_4}) increases, and a pronounced shoulder develops around 91 eV, and attains maximum intensity when the incident energy is tuned to the La $^3D_1$ resonance.

It is tempting to interpret this behavior as due to transitions involving La $4d^{-1}4f$ $^3D_1$   intermediate state. In this state the $f$ electron is localized at the lanthanum site, and in principle a valence electron can fill the core hole, to create a final state with a vacancy in the valence band and an excited $f$ electron. The sharp peak in the unoccupied La $f$ LPDOS of the electronic ground state (Fig.~\ref{fig:ElkSOCDOS}) suggests that the $f$ electron is localized also in the final state. According to dipole selection rules transitions from the valence band to a La $4d$ hole should be from state with La $p$ or $f$ character. However, the theory does not show appreciable LPDOS of these local symmetries (Fig.~\ref{fig:ElkSOCDOS}), and it lacks steep structures that could directly account for the observations.

%{\color{blue} Would it be possible to multiply the La $p$ and $f$ so that they are seen in fig. \ref{fig:ElkSOCDOS}?}
%They are visible

Tentatively, we still associate the resonant behavior to transitions to the La 4$d$ level from the valence band. With the excitation energy at 101.47 eV, and the emission energy of the resonant shoulder is 91 eV, the energy loss is 10.5 eV, suggesting that the corresponding hole is in the band that is dominated by Si $s$ states. Especially, the Si1 site, which is in closest proximity to the La atoms has a sharp peak at the corresponding energy (fig. \ref{fig:ElkSOCDOS}). The observations are consistent with a "cross transition" where the final state has an electron in the band dominated by La $f$-LPDOS and a vacancy in the band dominated by Si1 $s$-LPDOS, and we speculate that excitonic effects, and possibly interference between close-lying core-hole states may enable pathways to such final states.

Finally we note that there is a slight spectral change, when comparing data taken below and above the resonance. They are consistent with selective excitation of the two silicon sites, which show large variation at the cross section at these excitation energies (Fig. \ref{Figure_2}b).

\section{\label{sec:conclusion}Conclusions}

XAS and RIXS measurements at Si  $L$  and La $N$ edges of LaPt$_2$Si$_2$ are presented, and interpreted in terms of DFT calculations. Atomic-like local La $4d\rightarrow4f$ excitations are found in the absorption spectrum, the $LS$ forbidden $^3P_1$, and $^3D_1$ excitation as well as the giant $^1P_1$ resonance. Decay to $5p^{-1}4f$ final states are observed, and also the decay of the dynamically populated $^3D_1$ state to the electronic ground state. Observations suggest that resonantly excited $^3D_1$ states also decay via valence band emission. The Si $L$ XAS spectra are measured via PFY, and the XES spectra are  independent of the excitation energy over most of the energy range. The spectra are assigned using the theoretical LPDOS of Si $s$ and $d$ character at both crystallographic Si sites.

%To understand the LaPt2Si2 system, XAS and XES measurements were carried out. XAS shows La pre-resonances and giant resonance. While there are atomic like emission lines which shows the shifting of 4f levels as 4d core hole is created in La atom. Indirectly we can see that both Si1 and Si2 sites show different interacting properties and XES features. Most of these effects related to La atom are seen in other compounds of La, the specialty of this compound is that the core holes are very close to each other in energy this introducing new coupling within the atoms of Si and La.
	
%The measurements at room temperature shows electronic transitions in LaPt$_2$Si$_2$ that are unique. 

Since this system is similar to 122 Fe based superconductors and shows CDW formation at lower temperatures, we should expect to see 
%enhancement 
a variation in electronic interaction between La and 
%different 
Si atoms as a function of temperature, and an accompanying variation in the Si $L$ and La $N$ spectra.
%This would also account to how these Si and Pt interactions changes once CDW appears in the system. 
With site selectivity the question to what extent the CDW primarily is formed in the Si1-Pt2-Si1 or Pt1-Si2-Pt1 layers can be addressed. %are responsible for CDW in this system.

%The results are part of a  measurements were performed during a commissioning beamtime at the SPECIES beamline at MAX IV. 
A more detailed study will follow, including a systematic investigation of the temperature dependence across the CDW transition, and the dependence on crystal orientation.
	
\begin{ack}
The authors would like to thank Margit Andersson and Jenn-Min Lee for the technical support during the experiments at SPECIES beamline at MAX IV. Y.S. acknowledges funding from the Area of Advance - Material Sciences from the Chalmers University of Technology and a Swedish Science Council (VR) Starting Grant (Dnr. 2017-05078). M.M. and E.N. acknowledge funding from the Swedish Research Council (VR) through a neutron project grant (Dnr. 2016-06955), the Swedish Foundation for Strategic Research (SSF) within the Swedish national graduate school in neutron scattering (SwedNess), and the Carl Tryggers Foundation for Scientific Research (CTS-18:272).
%\textcolor{RubineRed}{Add number}(DNRxxxx).
M.A. acknowledges that SSF, Stiftelsen f{\"o}r strategisk forskning, through grant RIF14-0064 has provided funding. J.-E.R. acknowledges funding from the Swedish Science Council (VR) and the Carl Tryggers Foundation (CTS). The computations were enabled by resources provided by the Swedish National Infrastructure for Computing (SNIC) at PDC and NSC, partially funded by the Swedish Research Council through grant agreement no. 2018-05973.
\end{ack}

\section*{References}
\bibliographystyle{unsrt}
\bibliography{Ref}
	
\end{document}